\documentclass[amssymb,aps,twocolumn,floats]{revtex4}
\usepackage{color,graphicx,pstricks}

\begin{document}
\title{Frustration-induced insulating chiral spin state in itinerant triangular-lattice magnets}

\author{Sanjeev Kumar and Jeroen van den Brink}

\address{
Institute for Theoretical Solid State Physics, IFW Dresden, 01171 Dresden, Germany \\
}

\begin{abstract}
We study the double-exchange model at half-filling with competing superexchange interactions on a triangular lattice, 
combining exact diagonalization and Monte-Carlo methods. We find that in between the expected itinerant 
ferromagnetic and $120^{\circ}$ Yafet-Kittel phases a robust scalar-chiral, insulating spin state emerges. 
At finite temperatures the ferromagnet - scalar-chiral quantum critical point is characterized by 
anomalous bad-metal behavior in charge transport as observed in frustrated itinerant magnets 
R$_2$Mo$_{2}$O$_7$.
\end{abstract}

\date{\today} 


\maketitle

Geometric frustration is encountered in numerous magnetic condensed matter systems and gives rise to a 
variety of fascinating magnetic phases, such as spin-glass, spin-liquid, and 
spin-ice ~\cite{Gardner,Seabra-Stock}. Traditionally geometrically frustrated magnetism is studied in the 
context of wide bandgap materials with localized spins. Recently the interest broadened to frustrated 
magnets with metallic character such as Tl$_2$Mn$_2$O$_7$ and 
R$_2$Mo$_{2}$O$_7$  (R: rare earth ion) \cite{Shimakawa,Subramanian,Taguchi,Martinez}. In these metallic 
systems, frustration in the magnetic sector has a strong impact on charge dynamics, often resulting in 
heavy-fermion type behavior in transport 
\cite{Kondo,Iguchi,Harris,Bramwell}. 

Besides being of geometric origin, magnetic frustrations can also be due to a direct competition between 
ferromagnetic (FM) and antiferromagnetic (AFM) interactions. Perhaps the most elementary example is 
FM double-exchange (DE) competing with AFM 
superexchange (SE)~\cite{DeGennes1960}.  Magnetic competition of this kind has been studied extensively 
on hyper-cubic lattices in the context of  the colossal magnetoresistance manganites, where depending 
on carrier concentration the competing interactions can induce new forms of magnetic order or 
phase-separation~\cite{Dagotto_book,SK1,Brink1999}.  Competing FM and AFM interactions can in principle also lead 
to non-collinear and even non-coplanar magnetic states, which are raising interest in the rapidly 
emerging and seemingly disparate fields of multiferroics and topological 
insulators \cite{Khomskii,Martin,Akagi}.

In an interesting class of materials both types of frustrations -- geometric frustration 
{\it and} competing FM- and AFM interactions -- are present and give rise to a set of 
intriguing physical properties \cite{Ikoma,Motome10}. Recent experiments on pyrochlore 
R$_2$Mo$_{2}$O$_7$, for instance, show a transition from a FM metallic to a spin glass 
insulating state as R changes from Nd to Dy \cite{Iguchi}. This transition can be controlled by external 
pressure. In the vicinity of the transition an unusual diffusive metallic state appears, showing a 
temperature independent resistivity down to very low temperatures \cite{Iguchi}. A similar competition 
between FM and AFM interactions is also relevant in the  triangular lattice systems such as GdI$_2$ and 
H-doped GdI$_2$ \cite{Eremin,Deisenhofer,Ryazanov,Taraphder}. 
 
In this letter, we study in a wide temperature range the double-exchange (DE) model on a triangular lattice in the presence of frustrating AFM SE interactions. 
Three magnetically ordered groundstates are present in the phase diagram: an itinerant FM state on one end, the 120$^{\circ}$ Yafet-Kittel phase on the other and an insulating (I) non-coplanar scalar chiral (SC) state in between ~\cite{Yafet1952}. Apparently the SC-I state emerges from the competition between the two metallic phases: FM and 120$^{\circ}$. The FM and SC state are separated by a quantum critical point (QCP), whereas the transition from the SC to the 120$^{\circ}$ phase is first order in nature. At finite temperature in the vicinity of the QCP the system is characterized by a temperature-independent resistivity and a linear  inverse magnetic susceptibility, down to very low temperatures. This bad metallic behavior closely resembles the experimental observations on the R$_2$Mo$_{2}$O$_7$ compounds. 

We consider the elementary one-band DE Hamiltonian in the presence of antiferromagnetic SE interactions, introduced by De Gennes in the 1960's~\cite{DeGennes1960} -- but now on a frustrated triangular lattice. The full Hamiltonian is

\begin{eqnarray}
H &=& -\sum_{ \langle ij \rangle}
t_{ij} \left ( c^{\dagger}_{i} c^{~}_{j} + H.c. \right ) 
+ J_S \sum_{ \langle ij \rangle} {\bf S}_{i} \cdot {\bf S}_{j},
\end{eqnarray}

\noindent
where $c^{}_{i}$ and $c^{\dagger}_{i}$ are annihilation and creation 
operators for electrons with spin parallel to the core spin ${\bf S}_i$. 
$ \langle ij \rangle $ denotes the nearest neighbor (nn) pairs of sites on a triangular lattice. 
$J_S$ denotes the strength
of AF coupling between nn core spins.
$t_{ij} = t [\cos(\theta_i/2)\cos(\theta_j/2) + \sin(\theta_i/2)\sin(\theta_j/2) e^{{\rm -i} (\phi_i - \phi_j)}]$ 
denote the hopping amplitudes which depend on the polar and azimuthal angles 
\{$\theta_i$,$\phi_i$,$\theta_j$,$\phi_j$\} of the nn
core spins due to the double-exchange mechanism. All energies are measured in units of the hopping parameter $t$. The core spins are classical unit vectors and we focus on the case of half filling. 
The model Eq. (1) can be obtained as the large coupling limit of the Kondo lattice model (KLM),
$H_{KLM} = -t\sum_{\langle ij \rangle, \sigma} (c^{\dagger}_{i \sigma} c^{~}_{j \sigma} + H.c.) + J_K \sum_i 
{\bf S}_{i} \cdot {\bf \sigma}_{i}$. The 1st term in Eq. (1) corresponds to $|J_K| \rightarrow \infty$ in $H_{KLM}$, and the lowest order
correction ($\sim t^2/|J_K|$) leads to the superexchange term. 
Furthermore, the Hubbard model in the mean-field approximation has a similar structure as the KLM 
with the local moments originating self-consistently within a single band \cite{Martin}. Therefore,
the results discussed in this paper also bear relevance for the related models.

The only unbiased method to study this model is the hybrid scheme involving exact-diagonalization (ED) of the fermion problem and Monte Carlo (MC) for classical core spins \cite{Dagotto_book}. 
%The difficulty in handling this class of models and details on ED+MC method has been well documented in the context of manganites. 
In order to achieve larger lattice sizes, which are essential for computing transport properties, we employ the traveling cluster approximation (TCA) \cite{TCA}. This method is based on the observation that changes in total energy induced by a change in classical variable on a site can be estimated 
accurately by ED of a cluster Hamiltonian around that site.
The method has been benchmarked for similar models, and has proved successful in the study of manganites \cite{TCA,SK3}. Most of the results in this work are obtained on lattices with $N=24^2$ sites using a cluster size $N_c=6^2$. Typically $\sim 10^4$ MC steps are used for equilibration and a similar number of steps for computing thermal averages on classical spin variables. For electronic properties, which still requires the ED of the full Hamiltonian, $\sim 10^3$ steps are used.

\begin{figure}
\centerline{\includegraphics[width=.95\columnwidth,clip=true]{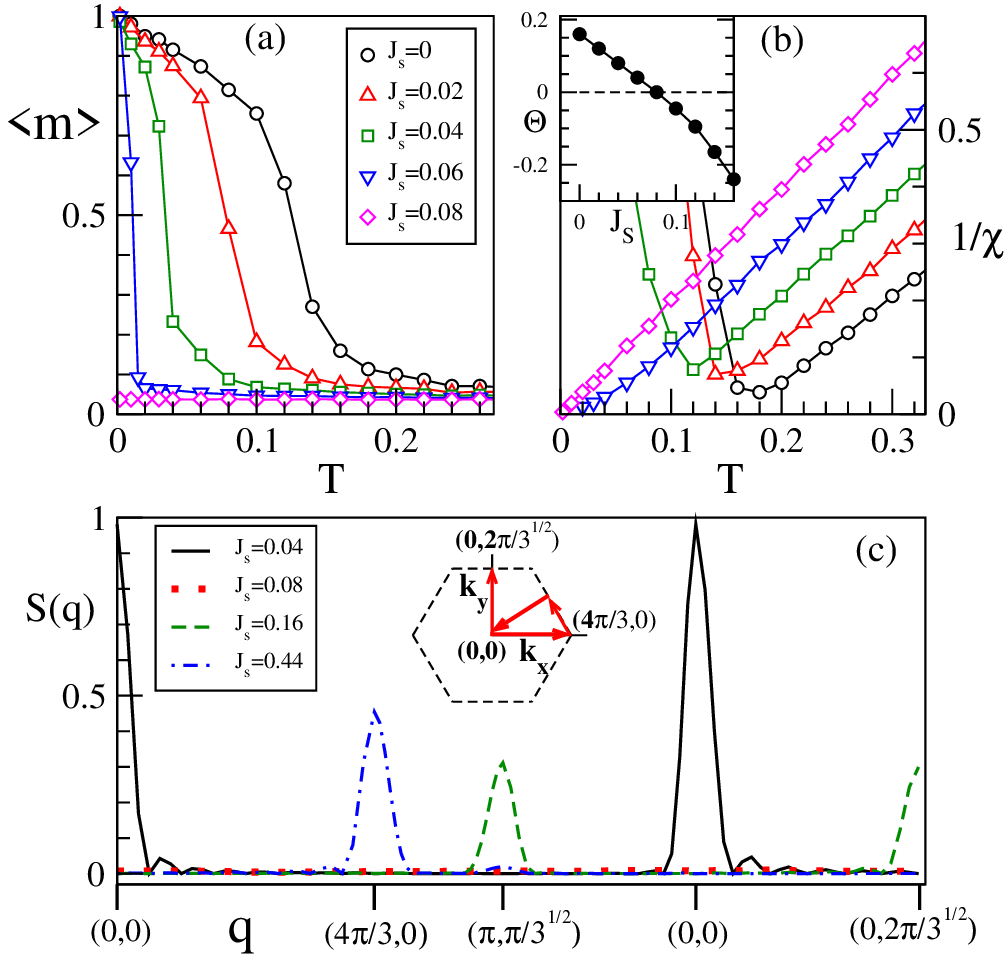}}
\caption{(Color online) (a) Magnetization and (b) inverse magnetic susceptibility as a function of temperature for various values of
superexchange coupling $J_S$. The inset in (b) shows the Curie-Weiss scale extracted from the $\chi^{-1}(T)$ as a function 
of $J_S$. (c) Spin-structure factor $S({\bf q})$ at
$T=0.001$ along the principal
symmetry directions in the momentum space for different values of $J_S$. The inset shows the Brillouin zone for a triangular lattice 
and the path along which $S({\bf q})$ is plotted.}
\end{figure}

%\vspace{0.4cm}

Fig. 1(a) shows the calculated temperature dependence of magnetization 
$ \langle m \rangle = \langle \left | 1/N\sum_i {\bf S}_i) \right |  \rangle$, where $\langle...\rangle$ denotes thermal averaging. In the absence of $J_S$ we find a FM groundstate with a Curie temperature $T_C = 0.15$ as inferred from the inflection point in the $\langle m \rangle (T)$ curve. Upon increasing $J_S$ a monotonic reduction in $T_C$ is observed while the groundstate continues to be a saturated FM. For $J_S = 0.08$ the ferromagnetism is destroyed by the competing AFM interactions and we do not find any long-range order in the groundstate. In Fig. 1(b) we show the inverse of magnetic susceptibility computed as $\chi = T^{-1}(\langle m^2 \rangle - \langle m \rangle^2)$. The inverse susceptibility $\chi^{-1}$ clearly follows a Curie-Weiss (CW) behavior  ($\chi \propto 1/(T-\Theta)$) above a characteristic temperature. The CW scale $\Theta$, which is obtained by  extrapolating the linear high-temperature behavior of $\chi^{-1}$, is shown in the inset in Fig. 1(b). $\Theta$ reduces monotonically and changes sign upon increasing $J_S$, indicating  a change in the nature of effective magnetic coupling from FM to AFM. 
A remarkable similarity of these results with those obtained on a pyrochlore lattice indicates that the essential physics is determined by the frustrated nature of the lattice \cite{Motome10}.

In order to identify non-trivial
long-range ordered magnetic phases we compute the spin structure factor,
$S({\bf q}) = 1/N^2 \sum \langle {\bf S}_i \cdot {\bf S}_j \rangle ~ e^{{\rm i} {\bf q} \cdot ({\bf r}_i-{\bf r}_j)}$, 
where ${\bf r}_i$, ${\bf r}_j$ are the real-space
location of spins ${\bf S}_i$, ${\bf S}_j$ . 
We show the groundstate structure factor along the symmetry directions in the momentum space in Fig. 1(c). 
For $J_S=0.04$, the $S({\bf q})$
has a single peak at ${\bf q} = (0,0)$ as expected for a FM state. In the regime $\Theta \sim0$ ($J_S=0.08$), the structure factor has 
no peak at any ${\bf q}$ suggesting the absence of any long-range magnetic order. For $J_S=0.16$, 
we find peaks at  ${\bf q} = (\pi,\pi/\sqrt{3})$ and $(0,2\pi/\sqrt{3})$. This corresponds to a four-sublattice ordered state with a 
non-vanishing scalar chirality $\kappa = \sum \langle {\bf S}_i \cdot {\bf S}_j \times {\bf S}_k \rangle$, 
where the sum is over indices forming a triangle taken in the counterclockwise order. The onset temperature $T_{sc}$ for
the SC state can be inferred from the temperature-dependence of $\kappa$. This state
has been discussed for the KLM within mean-field and variational schemes, and has also been studied in the context of
high-Tc superconductors and recently the topological insulators \cite{chiral_old,Martin,Akagi}. 
Here we show that the SC state is the 
groundstate for the model
Hamiltonian Eq. (1) over a window of phase space, and does not rely on Fermi surface nesting \cite{Akagi}.
Finally as the $J_S$ becomes large the system should approach towards the classical $120^{\circ}$ state. 
This is seen in the $S({\bf q})$ data for $J_S=0.44$, where the peak is located at ${\bf q} = (4\pi/3,0)$.

\begin{figure}
\centerline{
\includegraphics[width =.95\columnwidth , clip=true]{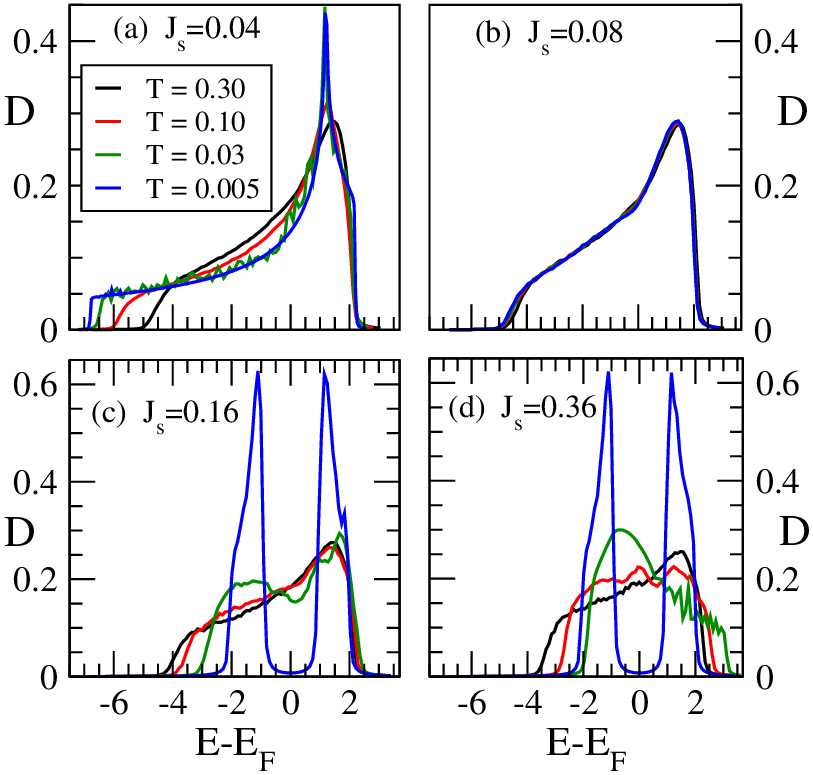}
}
\caption{(Color online) Electronic density of states for varying temperatures with (a) $J_S = 0.04$, (b) $J_S = 0.08$,
(c) $J_S = 0.16$ and (d) $J_S = 0.36$. A Lorentzian broadening $\gamma=0.04$ is used.
}
\end{figure}

%\vspace{0.4cm}

To investigate the repercussions of the competing magnetic interactions on electronic properties, we first focus on the evolution with $J_S$ of the electronic density of states (DOS), $D(E) = \langle \sum_k \delta(E-\epsilon_k) \rangle$, where we approximate the $\delta$-function by a Lorentzian of width $\gamma$: $\pi \delta(E-\epsilon_k) \sim {\gamma}/{[\gamma^2 + (E-\epsilon_k)^2]}$.
The DOS for different temperatures and values of $J_S$ is shown in Fig. 2. For small $J_S$ the DOS
does not show any drastic changes in shape upon reducing T (see Fig. 2(a)) and an increase in bandwidth is 
found as expected in a DE model. 
For $J_S = 0.08$ the DOS at low temperatures is identical to that in the high temperature paramagnetic (PM) state
(see Fig. 2(b)). This unusual behavior of the DOS is
consistent with the facts that (i) $\chi^{-1}$
follows the Curie-Weiss behavior down to $T=0$, and (ii) the spin-state remains disordered down to low temperatures
for $J_S=0.08$ (see Fig. 1). 
For $J_S = 0.16$, where the groundstate has the peculiar non-coplanar order, an energy gap
appears in the DOS near $E_F$ at low temperatures (see Fig. 2(c)). This is consistent with the previous discussions on this
non-trivial state \cite{chiral_old,Martin,Akagi}.
The effect of this energy gap persists at higher temperatures in the form of a pseudogap feature at $E_F$.
For larger $J_S$ an unusual flip in the DOS shape w.r.t. the high temperature DOS 
occurs near $T=0.03$ (see Fig. 2(d)) \cite{EPL_triangular}. 
We find that this
flip in the shape of DOS is correlated with the rise in $S({\bf q})$ at ${\bf q} = (4\pi/3,0)$ and hence indicates 
a $120^{\circ}$ state. 

We compute the dc conductivity ($\sigma(\omega)|_{\omega \rightarrow 0}$) using the Kubo-Greenwood formula and the exact
eigenspectrum.
In this work we use $\sigma(\omega_0) \vert_{\omega_0 = 0.03}$ as an approximation to 
dc conductivity except for low-temperature metallic states, where we make use of the 
conductivity sum rule \cite{SK_transport,Dagotto_RMP}. 
A typical double-exchange behavior in resistivity is found in cases where the groundstate is FM (see Fig. 3(a)). For $J_S = 0.08$,
which corresponds to a disordered magnetic groundstate as seen in $S({\bf q})$, we find an unusually flat resistivity. 
Further increase in $J_S$ leads to an insulating behavior ($d \rho/dT < 0$) at low temperatures with a diverging resistivity at 
$T\sim0.01$. For $J_S = 0.30$ one finds a sharp reduction in $\rho$ at a characteristic temperature associated with the rise 
in $S({\bf q})$ at ${\bf q} = ({4\pi/3,0})$. $d \rho/dT$ remains positive in the low temperature regime until the system enters the
scalar chiral insulating state near $T=0.005$. Eventually, beyond $J_S = 0.40$ the groundstate becomes the $120^{\circ}$-state. 

In Fig. 3(b) we show resistivity as a function of $J_S$ for low and high temperatures. This data can be compared with the pressure
dependence of resistivity reported in \cite{Iguchi}, since application of external pressure alter the $J_S/t$  ratio and tunes the 
system 
across the FM to PM transition. We find the resistivity at high temperature increases monotonically, whereas at low temperatures 
it shows
a non-monotonic behavior. These features qualitatively agree with the experimental observation  \cite{Iguchi}. We also plot the DOS at $E_F$ as a 
function of $J_S$ for low and high temperatures. Naively at low temperatures one expects the resistivity to be inversely related to 
the DOS at $E_F$, since only the states near Fermi level contribute. We indeed find
this behavior at low temperatures: the peak is $\rho$ corresponds to a dip in $D(E_F)$. 

\begin{figure}
\centerline{
\includegraphics[width =.95\columnwidth , clip=true]{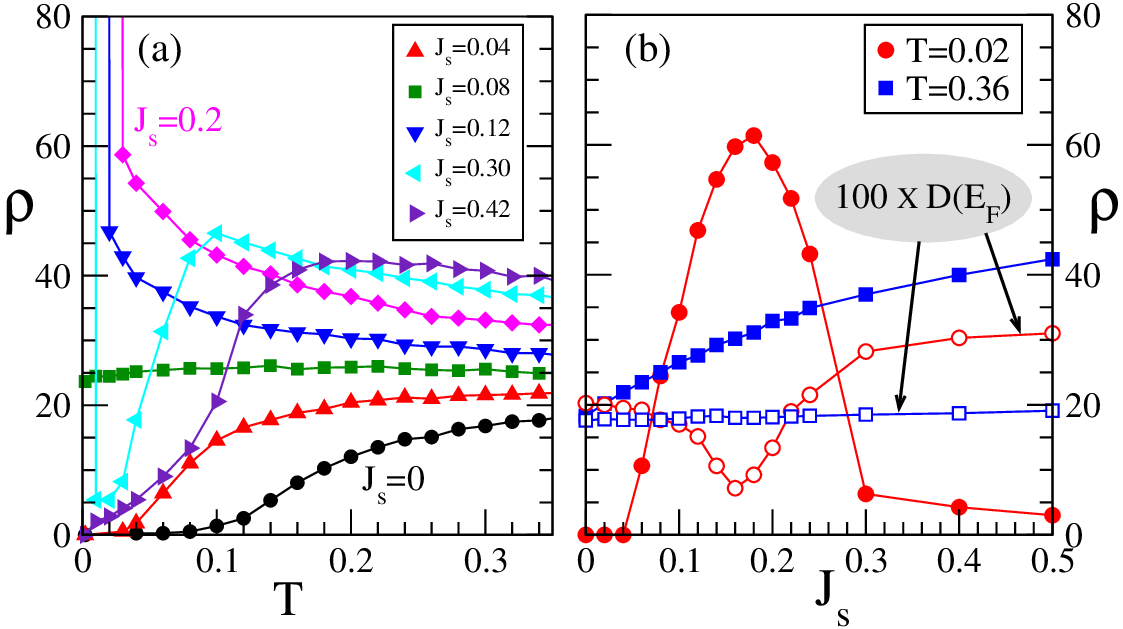}
}
\caption{(Color online) (a) Resistivity as a function of temperature for various values of $J_S$. (b) $J_S$-dependence of resistivity
$\rho$ (filled symbols) and scaled DOS (open symbols) at Fermi level $E_F$ for low and high temperatures.
}
\end{figure}

In Fig. 4(a) we plot the peak values in $S({\bf q})$ at various ${\bf q}$ at $T=0.005$ as a function of $J_S$.
Peak in $S({\bf q})$ at ${\bf q} = (0,0)$ is a measure for ferromagnetism, that at ${\bf q} = (4\pi/3,0)$ corresponds to $120^{\circ}$ state
and simultaneous peaks at ${\bf q} = (\pi,\pi/\sqrt{3})$ and $(0,2\pi/\sqrt{3})$ indicates a SC state. Upon increasing $J_S$,
$S(0,0)$ vanishes before 
a rise in $S(\pi,\pi/\sqrt{3})$ and $S(0,2\pi/\sqrt{3})$, indicating that the associated transition at $J_S \sim 0.07$ is quantum critical
in nature. On the other hand the transition between chiral and $120^{\circ}$ state is 1st order. The dashed curve shows the 
$J_S$-dependence of the scalar chirality 
$\kappa$.
The results are summarized in a phase 
diagram in Fig. 4(b).
The ferromagnetic transition temperature ($T_C$) is inferred from the 
$\langle m \rangle (T)$, the onset of pseudogap behavior is determined by directly looking at the DOS at various temperatures, and the transition to the
$120^{\circ}$ state is determined from the T-dependence of the $S({\bf q})$ at ${\bf q}=(4\pi/3,0)$. The absolute value $|\Theta|$
of the Curie-Weiss temperature is plotted as an estimate of the effective magnetic interaction in the system. The FM scale closely 
follows the value of $\Theta$, indicating the unfrustrated nature of the system in this regime. 
In the intermediate-$J_S$ regime the groundstate is SC-I and a pseudogap behavior appears just above $T_{sc}$.
This simple looking model provides an
elegant example where a competition between two metallic phases leads to an intermediate insulating phase.

\begin{figure}
\centerline{
\includegraphics[width =.85\columnwidth , clip=true]{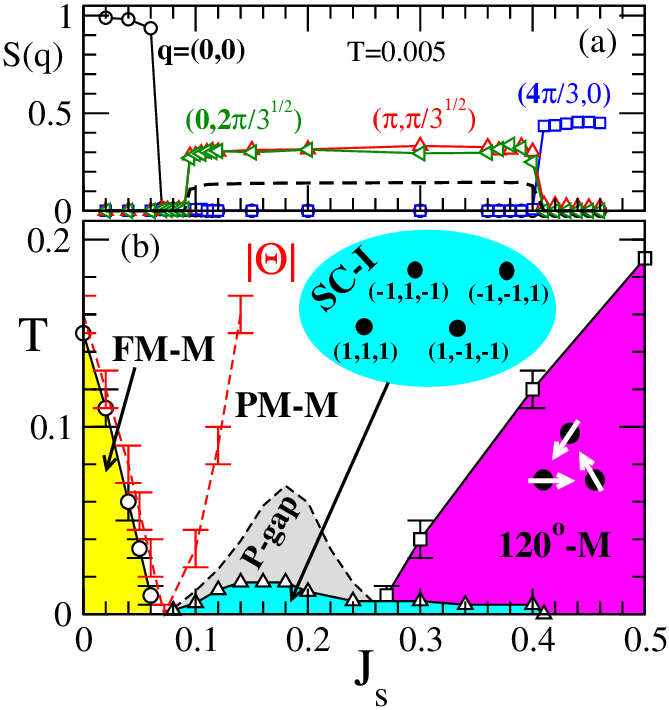}
}
\caption{(Color online) (a) Low-temperature spin structure factor $S({\bf q})$ as a function of superexchange coupling $J_S$ for various 
${\bf q}$. The dashed line is the scalar chirality $\kappa$. (b) The phase diagram 
of the model in the temperature-$J_S$ plane. The spin directions within a unit cell are schematically shown 
for the $120^{\circ}$ state, and are indicated as un-normalized ($S_x,S_y,S_z$) on each site for the non-coplanar SC state.
}
\end{figure}

For a simple physical picture of the various magnetic states
we derive an effective model for the DE term as $ H_{eff} = -\sum_{ \langle ij \rangle} K_{ij} cos(\alpha_{ij}/2)$,
with $cos(\alpha_{ij}) = {\bf S}_{i} \cdot {\bf S}_{j}$, and
$K_{ij} = t \langle e^{{\rm i} \Phi_{ij}}c^{\dagger}_{i} c^{~}_{j} + H.c. \rangle $, here the angular bracket denotes the
quantum expectation value in the groundstate \cite{DDE}. The phase $\Phi_{ij}$ in 
$ t_{ij} \equiv |t_{ij}| e^{{\rm i} \Phi_{ij}} $ is 
trivial for the FM and 120$^{\circ}$ states, which leads to real-valued hopping parameters. 
Minimizing $J_S cos(\alpha_{ij}) - K_{ij} cos(\alpha_{ij}/2)$ for a 
single bond with $K_{ij} = K_{FM} \sim 0.32 t$ we find that the 
FM state becomes unstable for $J_S > 0.25 K_{FM} = 0.08 t$, which is very close to the computed transition point.
In contrast, the SC-I state has non-trivial $\Phi_{ij}$, 
which give rise to local magnetic field terms in the
effective Hamiltonian.
The total phase around a closed plaquette 
forming the 4-site unit cell is
$\pi$, and a staggered sign of this flux between neighboring plaquettes explains 
the opening of energy gap in the spectrum of the SC state.
The self-consistent 
emergence of this magnetic flux clearly relies on the presence of itinerant fermions.
A similar chiral spin state has also been recently reported in a KLM 
on a pyrochore lattice \cite{Gia-Wei}. 
We also find non-trivial spin states for the present model on a checkerboard lattice \cite{SK-JvdB}, 
indicating that the results discussed here in detail for the triangular lattice have broad implications.

To conclude, a scalar-chiral insulating state emerges when double exchange and superexchange interactions 
compete on a triangular lattice. The transition from ferromagnetic metal to a scalar-chiral insulator is quantum critical in nature with both $T_C$ and $T_{sc}$ vanishing at a critical value of $J_S$. We find that
the chiral order is a consequence of a self-consistently generated magnetic flux, which relies crucially on
the presence of itinerant fermions. There is increasing experimental evidence that such chiral spin states 
lead to anomalous Hall effect via the hopping generated magnetic fields \cite{Taguchi,Taguchi-Machida}.
The transition from a FM-metal to a spin-glass insulating phase reported in the experiments on 
R$_2$Mo$_{2}$O$_7$ is qualitatively similar to that from FM-M to the SC-I phase \cite{Iguchi}. The
observed flat-resistivity in the vicinity of the transition is very well captured within our calculations.

SK acknowledges stimulating discussions with A. Mukherjee, G. V. Pai, S. R. Hassan and A. Paramekanti.

\end{document}